# Optimized Design of Survivable MPLS over Optical Transport Networks

Wojtek Bigos[1], Stéphane Gosselin[1], Morgane Le Foll[1], Hisao Nakajima[1], Bernard Cousin[2]

[1] France Télécom R&D – CORE/MCN, Lannion, France
E-mail: {wojtek.bigos, stephane.gosselin, morgane.lefoll, hisao.nakajima}@rd.francetelecom.com

[2] University of Rennes, IRISA Research Laboratory, Rennes, France
E-mail: Bernard.Cousin@irisa.fr

*Abstract* – **In this paper we study different options for the survivability implementation in MPLS over Optical Transport Networks (OTN) in terms of network resource usage and configuration cost. We investigate two approaches to the survivability deployment: single layer and multilayer survivability and present various methods for spare capacity allocation (SCA) to reroute disrupted traffic.**
**The comparative analysis shows the influence of the traffic granularity on the survivability cost: for high bandwidth LSPs, close to the optical channel capacity, the multilayer survivability outperforms the single layer one, whereas for low bandwidth LSPs the single layer survivability is more cost-efficient. For the multilayer survivability we demonstrate that by mapping efficiently the spare capacity of the MPLS layer onto the resources of the optical layer one can achieve up to 22% savings in the total configuration cost and up to 37% in the optical layer cost. Further savings (up to 9 %) in the wavelength use can be obtained with the *integrated* approach to network configuration over the *sequential* one, however, at the increase in the optimization problem complexity. These results are based on a cost model with actual technology pricing and were obtained for networks targeted to a nationwide coverage.**

I. INTRODUCTION

We consider the MPLS over Optical Transport Network (OTN) as a multilayer network architecture, where label switching routers (LSR) making up the *MPLS layer* are directly attached to optical cross-connects (OXC) belonging to the *optical layer*. In the optical layer, optical cross-connects are interconnected with point-to-point WDM links in a mesh topology. The interconnection between routers in this architecture is provided by circuit-switched, end-to-end optical channels or *lightpaths*. A lightpath (LP) represents a sequence of fiber links forming a path from a source to a destination router together with a single wavelength on each of these links. The OXCs can switch wavelengths between fiber links without undergoing optoelectronic conversion. A lightpath must be assigned the same wavelength on each link on its route, unless the OXCs support the wavelength conversion capability. The set of lightpaths established in the network makes up a *logical network topology*. IP traffic in the form of label switched paths (LSP) is carried in the network over this logical topology using a single or multiple *logical hops*. Fig. 1 shows an example of the MPLS over OTN.

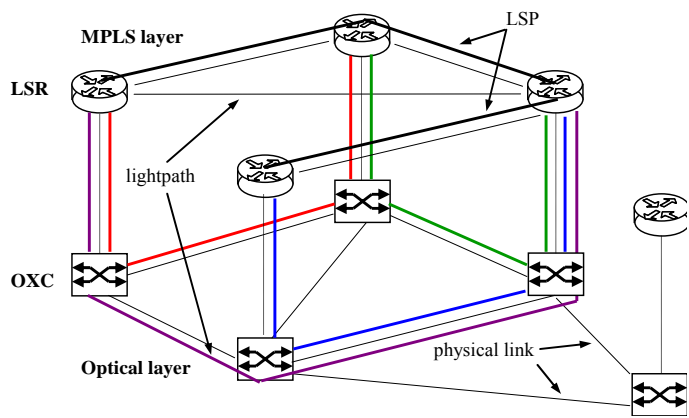

Fig. 1. Architecture of the MPLS over OTN.

As network survivability plays critical role in the network design, a number of recovery schemes have been proposed in the scope of the MPLS over OTN architecture. They are based on two general concepts: *single layer* survivability [1, 2], where recovery mechanisms are implemented only in the MPLS layer and *multilayer* survivability [1, 3, 4], where recovery is employed both in the MPLS and the optical layer. The multilayer survivability has the advantage over the single layer approach in faster and simpler recovery from physical link failures but it is considered to consume more optical layer resources [1, 4]. This is because with multilayer recovery each network layer reserves some spare resources for rerouting of affected paths, so multiple spare capacity pools are provided, each dedicated to a particular layer. On the other hand, single layer recovery requires more resources from the MPLS layer what may negatively affect the total network configuration cost, as these resources are more expensive than the resources of the optical layer.

In this paper we present the design of a survivable MPLS over Optical Transport Network (OTN) as an integer linear programming (ILP) optimization problem. Our objective is to minimize the amount of network resources used with a given network configuration. There are implemented different methods for spare capacity allocation (SCA) with the single- and multilayer survivability to reroute disrupted traffic. The planning process for SCA is based on two approaches to the MPLS over OTN configuration: the *sequential* one, where the MPLS layer and the optical layer are planned separately and the *integrated* one, where the whole network is designed in one step. The aspects of spare capacity planning and sequential/integrated approaches to network configuration are related to the design of *multilayer* network architectures and contribute to efficient network configuration in terms of resources usage. The objective of this work is to consider both these aspects in the context of a network optimization problem and to investigate their impact on network resource savings. A set of MPLS over OTN configurations is implemented, where a particular SCA method is combined with a particular configuration approach demonstrating their relative importance to the overall network design in terms of network resource consumption and configuration cost.

The rest of this paper is organized as follows: Section II presents different SCA methods for the single- and multilayer survivability implementation. Section III describes two approaches to the MPLS over OTN configuration: sequential and integrated and explains their impact on the network resource usage. We precise the objectives to be realized with respect to these problems and explain how our work extends the previous studies. In Section IV we define a framework for the survivable MPLS over OTN design. We present algorithms for different spare capacity planning options, define a cost model to be included into the optimization procedure and give exact ILP formulations for the considered problems. Section V concludes with the analysis of the obtained results.

## II. SPARE CAPACITY PLANNING IN MPLS OVER OTN

One of the aspects related to the survivability design is how to allocate spare capacity in a network, so that the total amount of network resources is minimized. The total amount of network resources used with a given SCA option depends on supported failure scenarios and a recovery technique used. In this study we employ *end to end path protection* as a recovery technique in individual network layers; it is believed that such a recovery scheme will be always required in a network to protect the integrity of high class services. The considered failure scenarios include physical link failures (e.g. fiber cuts, optical line system failures), *transit* [1] node failures (both router and OXC) and IP/optical interface failures.

With the single layer survivability (Fig. 2a) protection is implemented at the LSP level and the MPLS layer covers all failure types. Each *working* LSP (wLSP) has a corresponding *protection* LSP (pLSP), link- and node-disjoint in both network layers. With the multilayer survivability (Fig. 2b) protection is implemented both at the LSP level (pLSP) and the lightpath level (pLP). The optical layer protects against physical link and OXC failures, whereas the MPLS layer protects against router and IP/optical interface failures which cannot be detected by the optical layer [2]. This implies that, with the multilayer survivability, only *multi-hop* LSPs which are susceptible to router failures have corresponding protection LSPs routed in the MPLS layer (to provide node disjointness against router failures). *Single-hop* LSPs do not require any extra spare capacity from the MPLS layer as they are subject only to the failures resolved at the lightpath level (e.g. wLSP2 carried on wLP4 is protected by pLP4). To cover the IP/optical interface failures (e.g. optical line cards, intra-office links and tributary OXC ports), the reach of protection lightpaths is extended towards optical line cards in routers.

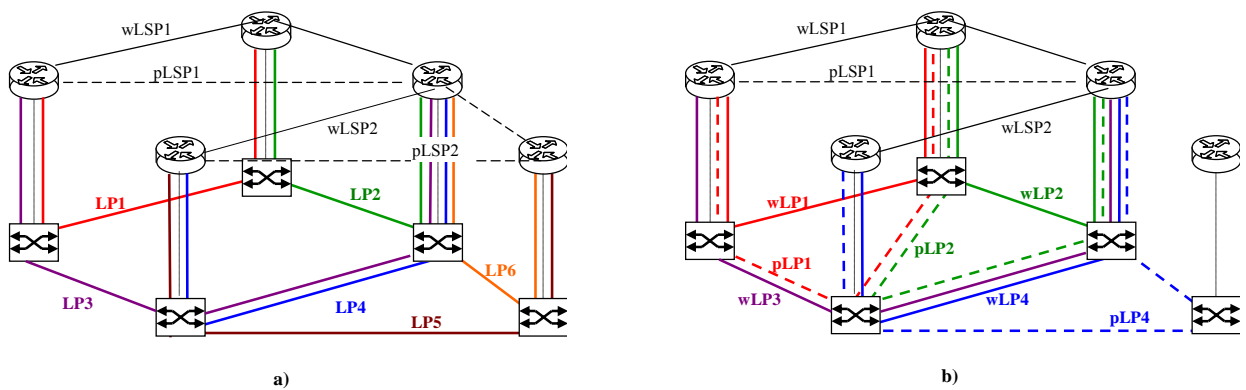

Fig. 2. Two options for the survivability implementation in MPLS over OTN: a) single layer survivability, b) multilayer survivability.

---

[1] Paths originating/terminating at a failed node are considered as lost, since they cannot be restored with path protection mechanisms.

Another point to consider with the multilayer survivability implementation is how the MPLS spare capacity used to protect working LSPs is supported by the optical layer. With a simple capacity planning without any precautions taken, called *double* or *redundant protection* [2], spare capacity in the MPLS layer is protected again in the optical layer. The working LSPs are thus twice protected: once in the MPLS layer and once in the optical layer. This results in inefficient use of network resources with very little increase in service reliability. An improvement in the optical spare capacity utilization can be achieved by supporting working and protection LSPs on different lightpaths and treating them differently in the optical layer: lightpaths carrying working LSPs are protected while lightpaths supporting protection LSPs are left unprotected (e.g. wLP3 in Fig. 1b carrying protection LSP pLSP1). This option, called *LSP 'spare' unprotected* [2] requires less resources than double protection; it is however still inefficient in a way that the optical layer still dedicates some resources to support the MPLS spare capacity. Further improvement in spare capacity planning consists in sharing spare resources between the MPLS and the optical layer. With this option, called *interlayer backup resources sharing* (interlayer BRS) [3] or *common pool survivability* [1], [4] the MPLS spare capacity is considered as extra traffic in the optical layer (i.e., carried on unprotected, pre-emptible lightpaths, such as wLP3 in Fig. 1b). As a consequence, there exists only one spare capacity pool (in the optical layer, for lightpath protection) which can be reused by MPLS recovery schemes when needed (e.g. in Fig. 2b protection lightpaths pLP1 and pLP2 share the wavelengths with the lightpath wLP3 carrying protection LSP pLSP1). Little or no additional resources in the optical layer are required to support the MPLS spare capacity.

The analysis of the survivability implementation in the MPLS over OTN in terms of resources usage and configuration cost has been already provided in the literature. The analysis provided in [5] shows that the single layer recovery is by 10% more cost-effective than the multilayer recovery when lightpaths are not fully utilized with the working traffic, whereas for the high lightpath utilization it is the opposite. The authors however consider only physical link failures as a possible failure scenario (assuming dual-router architecture to protect against router failures) and the results taking into account more failure scenarios may be different. Various SCA options for the multilayer survivability has been investigated in [3] showing 15% and 20% cost improvements achieved respectively with the "LSP spare unprotected" and the "interlayer BRS" methods over the "double protection". The planning process used in [4] does not guarantee the recovery from the OXC failures when using the interlayer BRS option. This work extends the previous studies by adding more failure scenarios, including physical link failures, node failures (both router and OXC) and IP/optical interface failures. The exact planning processes are given for different SCA options and a cost model is defined which allows the network configuration cost to be modified according to the price evolution of network components. Finally, we present the integrated approach to the network design where the MPLS and optical layer are optimized jointly, leading to extra savings in network resources.

III. SEQUENTIAL VS. INTEGRATED APPROACH TO THE MPLS OVER OTN CONFIGURATION

In the MPLS over OTN both network layers can be combined using either the *overlay* or the *peer* interconnection model [6]. In the overlay model, the MPLS and the optical layer are controlled separately and each layer has its own instance of the control plane. There are two separate routing processes in the network: the MPLS layer routes LSPs in the logical topology using either existing lightpaths or requesting a new lightpath *directly* connecting LSP endpoints from the optical layer; then the optical layer routes the lightpaths in the physical topology. In the peer model, a single control plane controls both the MPLS and the optical layer. There is one routing process which runs across both layers and logical and physical links are considered jointly in route selection.

Most previous studies on the routing implementation either with the overlay [2] or the peer interconnection model [7]-[9] concentrate exclusively on analyzing the blocking probability of a given routing approach under dynamic traffic conditions. There are however no analyses showing their performance in terms of network resources consumption. Our objective is to implement the concepts of sequential and integrated routing as network optimization problems and to compare the use of network resources achieved with both methods. The considered network resources include:

- In the MPLS layer:
- amount of packet processing in routers which is proportional to the volume of transit traffic at each router (i.e. neither originating nor terminating in a router); by minimizing it we improve the router throughput and minimize the

---

[2] The MPLS layer also protects against OXC failures with respect to the LSPs transit in co-located routers.

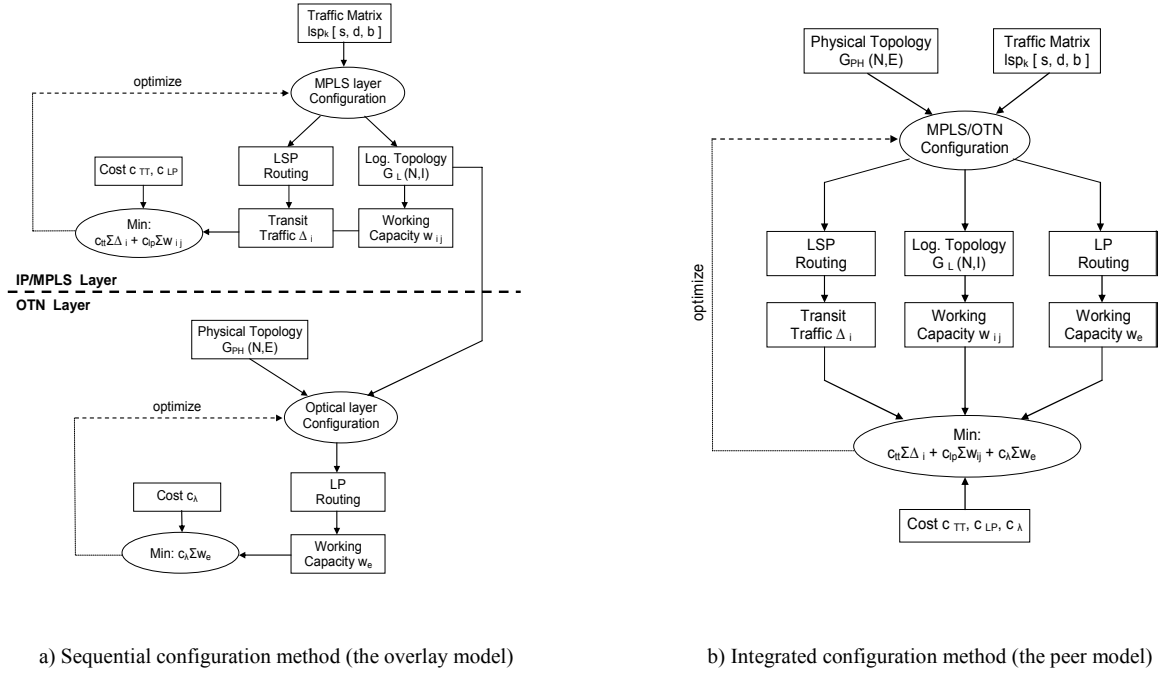

a) Sequential configuration method (the overlay model)  b) Integrated configuration method (the peer model)

Legend:
traffic matrix – set of K LSPs; each LSP has associated (s)ource, (d)estination point and (b)andwidth; $G_{PH}(N,E)$ – graph representing the physical network topology of N vertices (nodes) and E edges (physical links); $G_L(N,I)$ - graph representing the logical topology of N nodes and I logical links; $w_{i,j}$ –capacity per node pair in the MPLS layer, measured in the number of lightpaths installed between the node pair i-j; $\Delta_i$ – transit traffic processed by the node i; $w_e$ –capacity per link in the optical layer, measured in the number of wavelengths used on the physical link e to carry lightpaths; $c_{LP}$ – the lightpath cost; $c_\lambda$ – the wavelength cost; $c_{TT}$ – the transit traffic cost.

Fig. 3. Sequential vs. integrated approach to the MPLS over OTN configuration

- packet queuing delay.
- number of IP/optical interfaces in routers which constitute a significant part of the configuration cost;

- In the optical layer:

- number of wavelengths and wavelength-switching equipment used to route a given set of lightpaths in the physical topology.

With the sequential approach to network configuration, as presented in Fig 3a, the optimization problem consists of two sub-problems. First sub-problem takes as an input a traffic matrix in terms of the LSPs to be routed in the network and returns as a result the set of lightpaths to be established in the optical layer (i.e. the logical topology) and the routing of the LSPs over the logical topology. Thus, only a part of the MPLS over OTN configuration is solved by the first sub-problem and the optimization function takes as an objective minimizing *only* the IP/MPLS layer resources, i.e. the total number of lightpaths between all node pairs $\sum_{(i,j)} w_{(i,j)}$ and the total transit traffic $\sum_i \Delta_i$ processed by routers. Second sub-problem takes the set of lightpaths to be established on physical links and the physical network topology as input parameters and returns the routing of lightpaths in the physical topology optimizing *only* resources of the optical layer, i.e. the total number of wavelengths $\sum_e w_e$ used to route the lightpaths in the physical topology. The lightpath cost $c_{LP}$, the wavelength link cost $c_\lambda$ and the transit traffic cost $c_{TT}$ are included into the optimization procedure to account for the total configuration cost of the network. Note, however, that such decomposition is approximate or inexact. Solving the sub-problems in sequence and combining the solutions may not result in the optimal solution for the fully integrated problem. On the other hand, with the integrated approach, as depicted in Fig. 3b, there is only one optimization problem implemented which provides a full MPLS over OTN design in one step. The optimization function used to control the configuration process explicitly accounts for the resource usage both in the MPLS and the optical layer. The global optimization of network resources is thus possible with the integrated approach. Nevertheless, achieving an absolutely optimal solution of such a problem may be hard as the computational complexity of the algorithm solving the network configuration problem in a combined fashion is high.

IV. Problem Formulation

We consider the following network design problem. Given the offered traffic matrix (in terms of static LSP connections), the physical network topology and a set of constraints on logical and physical link capacities, we search for an MPLS over OTN configuration which (i) provides 100% restorability against the considered failure scenarios and (ii) minimizes the total resource usage in both network layers. A cost model is included into the optimization procedure which represents the monetary cost of various network components. Thereby, by minimizing the amount of network resources the cost of a given network configuration is also optimized. We use integer linear programming (ILP) as an optimization technique and formulate the problem using linear models. The problem solution provides a complete specification to the logical topology design and routing of working and protection paths both in the MPLS and the optical layer together with the resource usage at the minimal cost.

We make the following assumptions in our study:
1) The traffic matrix is symmetric and the lightpaths are bidirectional. Two lightpaths from a pair are routed over the same physical route but in the opposite directions.
2) The optical layer has an *opaque* configuration with photonic OXCs (i.e. which switch wavelengths optically) surrounded by WDM transponders performing OEO conversion. Transponders perform signal regeneration and adaptation functions including wavelength conversion.
3) As the optical layer supports wavelength conversion, the wavelength continuity constraint is not considered (under which a lightpath is assigned the same wavelength on all links on its route). This assumption reduces the problem in terms of ILP variables and constraints and makes it more computationally tractable.

*A. Survivability Implementation*

Algorithms for the MPLS over OTN design using the *sequential* configuration approach combined with various options for the survivability implementation are presented in Fig. 4 a-d. Each algorithm consists of four planning processes (steps): two for the network design with *working* paths, respectively in the MPLS and the optical layer (steps I and III), and two others for the network configuration with *protection* paths (steps II and IV). Each part is implemented as a separate optimization problem using a distinct ILP formulation.

The planning process for the computation of working LSPs and lightpaths (i.e. steps I and III in each algorithm) is analogous to the one presented in Fig. 3a. Next, the routing of protection paths is determined for the set of pre-computed working paths and spare capacity is allocated. With the single layer survivability (see Fig. 4a), the spare capacity planning is done only in the MPLS layer (in step II). The planning process takes as the inputs the considered failure scenarios and the working LSP routing computed in step I. Then, the routing of protection LSPs is performed analogously to the working LSPs routing, but respecting the constraints for protection routing in the MPLS layer (see below). The optimization function aims at minimizing the total transit traffic $\sum_i \Delta_i$ and the spare capacity $\sum_{(i,j)} s_{(i,j)}$ consisting of the lightpaths carrying protection LSPs. With the multilayer survivability implementation (see Fig. 4b-d), the spare capacity planning is done both in the MPLS and the optical layer with the objective to minimize the resources of each layer. Contrary to the single layer survivability, only multi-hop LSPs are subject to the protection LSP routing in step II; single hop LSPs are protected at the lightpath level. With the "double protection" method (Fig. 4b), both the lightpaths carrying working and protection LSPs are protected in the optical layer. With the "LSP spare unprotected" option (Fig. 4c), working and protection LSPs are routed over two disjoint sets of lightpaths (sets $G_{L1}$ and $G_{L2}$) and only the lightpaths carrying working LSPs (from $G_{L1}$) have protection lightpaths; the lightpaths carrying protection LSPs (from $G_{L2}$) are not protected. With the "interlayer BRS" (Fig. 4d), not only the lightpaths carrying protection LSPs are unprotected but they also share the wavelengths with the protection lightpaths to optimize further the wavelength use (i.e. spare capacity $s_e$ is shared with the working capacity $w_{e2}$ supporting the lightpaths which carry protection LSPs).

Next, the planning process was repeated but using the integrated configuration approach where the routing of working LSPs and the corresponding lightpaths (i.e. steps I, III) was implemented as a single optimization problem according to the scheme presented in Fig 3b. The same methodology was then used for protection LSPs and the lightpaths carrying protection LSPs (steps II and III). Note, however, that to combine the routing of working and protection LSPs and lightpaths (i.e. steps I – II and III – IV) we use the sequential approach only, i.e. routing of protection paths is determined for the set of pre-computed working paths. The results from computing working and protection paths jointly are reported in [10] showing 8 % - 12 % savings in spare capacity than if the paths are computed separately, however, at the increase in the optimization problem complexity.

The following rules are defined for the protection routing, i.e., how to route protection paths, so that all the considered failure scenarios are supported:
I. For the single layer survivability:

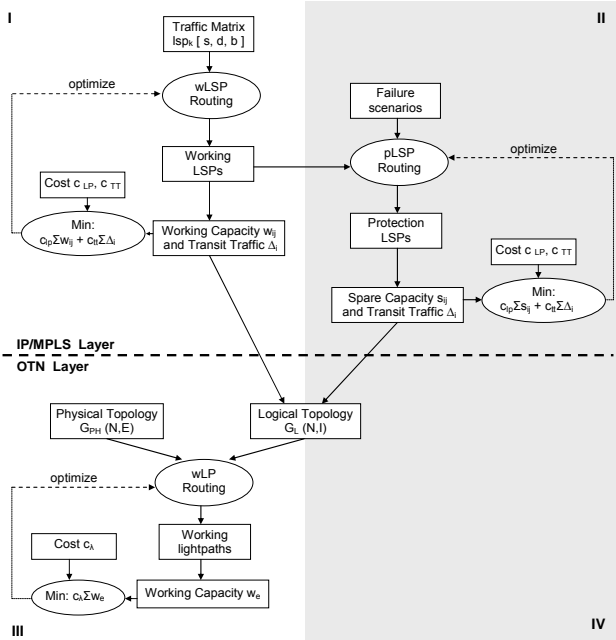

a) Single layer survivability

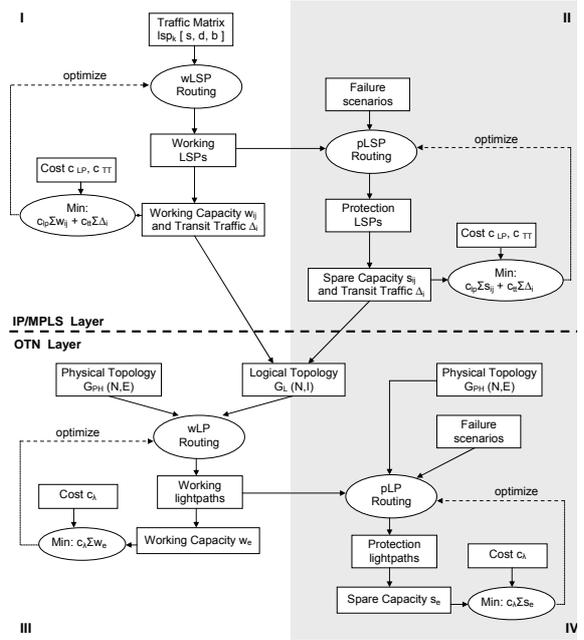

b) Multilayer survivability with double protection

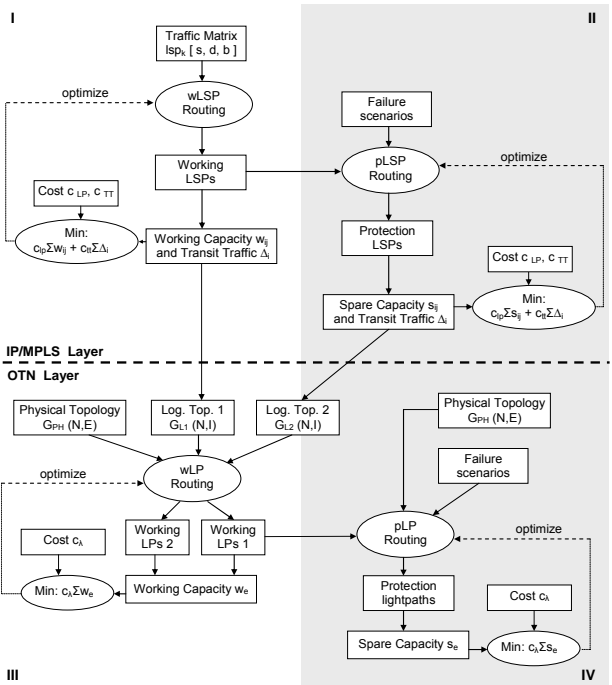

c) Multilayer survivability with the "LSP spare unprotected" option

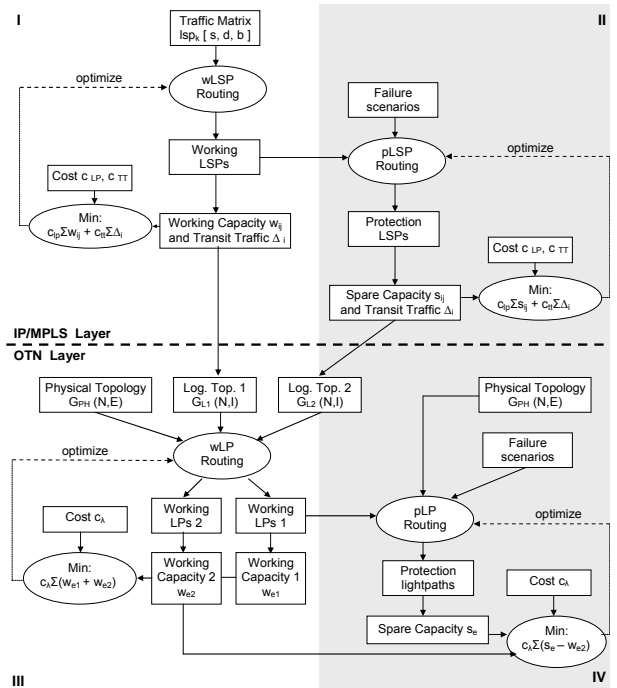

d) Multilayer survivability with the interlayer BRS

Legend:
$s_{i,j}$ – spare capacity per node pair in the MPLS layer, measured in the number of lightpaths carrying protection LSPs (pLSP) between the node pair i-j;
$s_e$ - spare capacity in the optical layer, measured in the number of wavelengths used on the physical link e to carry protection lightpaths (pLP); The other symbols have the meaning as specified in Fig. 2

Fig. 4. Algorithms for the MPLS over OTN design using the sequential approach with various options for the survivability implementation.

1. Each working LSP (wLSP) has a protection LSP (pLSP); corresponding working and protection LSPs are node- and link disjoint both in the in the logical and physical topology.
II. For the multilayer survivability

1. Each *multihop* working LSP has a protection LSP; corresponding working and protection LSPs are link- and node-disjoint in the logical topology.

2. With the "double protection" method each lightpath has a protection lightpath, with the "LSP spare unprotected" and "interlayer BRS" method only the lightpaths carrying working LSPs have corresponding protection lightpaths. Respective working and protection lightpaths are link- and node-disjoint in the physical topology.

Additional requirement for the "LSP spare unprotected" and "interlayer BRS" methods:

3. Corresponding working and protection LSPs are node disjoint in the physical topology; this prevents a working LSP and its protection LSP from failing simultaneously in case of an OXC failure.

And for the "interlayer BRS":

4. Lightpaths transiting an OXC and LSPs transiting the co-located router are protected on different physical links; this prevents the failed entities from competing for the same spare resources in case of an OXC failure.

### B. Cost Model

By including the cost of various network elements into the optimization procedure we optimize the network configuration cost which depends on the number and type of established communication channels. In this study we do not deal with the investments for the initial network deployment which include the cost of lying/leasing fibers and the cost of WDM line systems (without transponders), i.e. WDM (de)multiplexers and optical amplifiers. As it has been stated in the problem formulation, the initial network topology determining these costs is given to the problem as an input parameter.

The following cost components are included into the optimization procedure: the cost of IP/optical interface cards $c_{P\_IP}$, the cost of OXC ports $c_{P\_OXC}$ and the cost of optical transponders $c_{TR}$. The costs of the router and OXC equipment are incorporated, respectively, into the IP/optical interface cost $c_{P\_IP}$ and the OXC port cost $c_{P\_OXC}$. Additional cost $c_{TT}$ is associated with the amount of transit traffic processed by the routers as a penalty for diminishing the packet processing capability of a router which could be otherwise used by the originating/terminating traffic. The cost $c_{TT}$ is specified by the $c_{P\_IP}$ cost per traffic unit:

$$c_{TT} = \frac{c_{P\_IP}}{C} \times \text{transit traffic} \quad (1)$$

where C denotes the IP/optical interface card rate.

We take as a reference cost the cost of one transponder. The other elements' costs are referenced to this cost as follows:

$$c_{P\_IP} : c_{P\_OXC} : c_{TR} = 8 : 0.5 : 1 \quad (2)$$

This ratio represents the current prices of the elements (year 2005) for 10 Gbps IP/optical interfaces and transponders, 256×256 port *photonic* OXCs and 200 Gbps routers.

The total cost of a network configuration is a sum of the transit traffic cost and the cost of individual circuits: lightpaths and wavelength links, as depicted in Fig. 5. It is assumed that each wavelength link requires 2 transponders and 2 OXC ports (i.e. 1 transponder and 1 OXC port for each termination point) and each lightpath (protection and working) requires 2 IP/optical interface cards and 2 OXC ports. Thus, the cost components $c_{LP}$ and $c_\lambda$ presented in schemes 4 a-d are as follows: $c_{LP} = 2(c_{P\_IP} + c_{P\_OXC})$, $c_\lambda = 2(c_{P\_OXC} + c_{TR})$.

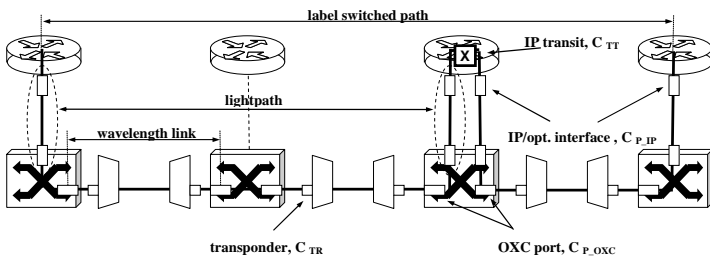

Fig. 5. An example of the MPLS over OTN configuration combined with the network elements' cost.

### C. ILP formulation

The ILPs are defined using the *node-arc* formulation, where network nodes are indexed by subscripts and edges are specified by an *(x,y)* node-name pair:

- *s* and *d* denote source and destination nodes of an LSP,
- *i* and *j* denote originating and terminating nodes in a lightpath,
- *q* denotes a *q-th* lightpath between the *(i,j)* pair,
- *m* and *n* denote endpoints of a physical link.

*Inputs:*

N      set of nodes; each element represents a generic network node being a combination of a router and the co-located OXC.

LSP      set of K LSPs to be routed in the network; each element $lsp_k$ represents an indivisible traffic flow to be routed on a single LSP and has associated a triple: {s($lsp_k$), d($lsp_k$), b($lsp_k$)} denoting respectively its source, destination node and bandwidth.

LP      set of lightpaths determining the logical topology; each element *lp* has associated a triple: {i(*lp*), j(*lp*), q(*lp*)} denoting respectively its originating, terminating node and the multiple.

p_LSP      set of LSPs to be protected; a subset of LSP.

p_LP      set of lightpaths to be protected; a subset of LP.

N_ex [*lsp*]      for each *lsp* in p_LSP, the set of nodes to be excluded from the route of its protection LSP to respect the protection routing constraints.

N_ex [*lp*]      for each lightpath in LP, the set of nodes to be excluded from its route to respect the protection routing constraints. Note, that in general not only protection but also working ligthtpaths are subject to the protection routing, e.g. to provide disjointness between (corresponding) working and protection LSPs in the physical topology with the single layer survivability.

$P_{NxN}$      physical topology matrix, where the element $p_{(m,n)} = p_{(n,m)} = 1$ if there exist a physical link between nodes *m* and *n*; otherwise $p_{(m,n)} = p_{(n,m)} = 0$, (i.e. physical links are bidirectional). It is assumed that there are no multiple links between node pairs.

C      lightpath capacity.

T      max. number of IP/optical interfaces in a router.

W      max. number of wavelengths on a physical link.

*Variables:*

1) Logical topology variables $w\beta_{(i,j),q}$, $p\beta_{(i,j),q} = 1$ if there exist a *q-th* lightpath from node *i* to node *j* carrying, respectively, working and protection LSPs; 0 otherwise.

2) LSP routing variables $w\delta^{lsp}_{(i,j),q}$, $p\delta^{lsp}_{(i,j),q} = 1$ if, respectively, the working / protection LSP *lsp* is routed on the *q-th* lightpath from node *i* to node *j*; 0 otherwise.

3) Lightpath routing variables:

   a) With the sequential configuration method: $w\lambda^{lp}_{(m,n)}$, $p\lambda^{lp}_{(m,n)} = 1$ if, respectively, the working / protection lightpath *lp* is routed on the physical link *(m,n)*; 0 otherwise.

   b) With the integrated configuration method: $w\lambda^{(i,j),q}_{(m,n)}$ if a *q-th* working lightpath from node *i* to node *j* is routed on the physical link *(m,n)*; 0 otherwise. Note, that as protection lightpaths are routed using the sequential method only, no protection routing variable is defined here.

*Optimization function:*

The objective is to limit the total network resource usage by minimizing costs of the transit traffic and capacity allocation (working + spare) in both network layers:

Minimize: $$c_{TT} \cdot \sum_n \Delta_n + c_{LP} \cdot \sum_{(i,j)} \left( w_{(i,j)} + s_{(i,j)} \right) + c_\lambda \cdot \sum_e (w_e + s_e) \qquad (3)$$

where:

$$\Delta_n = \sum_{lsp} b\_req(lsp) \cdot \sum_i \sum_q \left( w\delta^{lsp}_{(i,n),q} + p\delta^{lsp}_{(i,n),q} \right) - \sum_{lsp\,:\,d(lsp)=n} b\_req(lsp) \qquad (4)$$

$$w_{(i,j)} + s_{(i,j)} = \sum_q w\beta_{(i,j),q} + \sum_q p\beta_{(i,j),q} \tag{5}$$

- With the sequential configuration approach:

$$w_e + s_e = \sum_{lp} w\lambda^{lp}_{(m,n)} + \sum_{lp} p\lambda^{lp}_{(m,n)} \qquad \forall\ m, n : e = (m,n) \tag{6a}$$

- With the integrated configuration approach:

$$w_e + s_e = \sum_{(i,j)}\sum_q w\lambda^{(i,j),q}_{(m,n)} + \sum_{lp} p\lambda^{lp}_{(m,n)} \qquad \forall\ m, n : e = (m,n) \tag{6b}$$

The cost components $c_{TT}$, $c_{LP}$ and $c_\lambda$ are associated as specified in the cost model (see Sec. IV B)

*Constraints:*

1) Constraints for the logical topology design:

$$\sum_{j:j\neq i}\sum_q w\beta_{(i,j),q} + p\beta_{(i,j),q} \leq T, \qquad \forall\ i \in N \tag{7}$$

$$\sum_{i:i\neq j}\sum_q w\beta_{(i,j),q} + p\beta_{(i,j),q} \leq T, \qquad \forall\ j \in N \tag{8}$$

2) Constraints for the LSP routing:

$$\sum_{j:j\neq i}\sum_q w\delta^{lsp}_{(i,j),q} - \sum_{j:j\neq i}\sum_q w\delta^{lsp}_{(j,i),q} = \begin{cases} 1, & \text{if } i = s\,(lsp) \\ -1, & \text{if } i = d\,(lsp) \\ 0, & \text{otherwise} \end{cases} \qquad \forall\ lsp \in LSP,\ i \in N \tag{9}$$

$$\sum_{\substack{j:j\neq i,\\ j\notin N\_ex[lsp]}}\sum_q p\delta^{lsp}_{(i,j),q} - \sum_j\sum_q p\delta^{lsp}_{(j,i),q} = \begin{cases} 1, & \text{if } i = s\,(lsp) \\ -1, & \text{if } i = d\,(lsp) \\ 0, & \text{otherwise} \end{cases} \qquad \forall\ lsp \in p\_LSP,\ i \notin N\_ex[lsp] \tag{10}$$

$$w\delta^{lsp}_{(i,j),q} + p\delta^{lsp}_{(i,j),q} \leq 1 \qquad \forall\ lsp \in p\_LSP,\ (i,j) \in N^2,\ q \tag{11}$$

$$\sum_{lsp} b\_req\,(lsp) \cdot w\delta^{lsp}_{(i,j),q} \leq w\beta_{(i,j),q} \cdot C \qquad \forall(i,j) \in N^2,\ q \tag{12}$$

$$\sum_{lsp} b\_req\,(lsp) \cdot p\delta^{lsp}_{(i,j),q} \leq p\beta_{(i,j),q} \cdot C \qquad \forall(i,j) \in N^2,\ q \tag{13}$$

3) Constraints for the lightpath routing:

- With the sequential configuration approach:

$$\sum_{\substack{m:m\neq n,\\ m\notin N\_ex[lp]}} w\lambda^{lp}_{(n,m)} - \sum_m w\lambda^{lp}_{(m,n)} = \begin{cases} 1, & \text{if } n = i\,(lp) \\ -1, & \text{if } n = j\,(lp) \\ 0, & \text{otherwise} \end{cases} \qquad \forall\ lp \in LP;\ n \notin N\_ex[lp] \tag{14a}$$

$$\sum_{\substack{m:m\neq n,\\ m\notin N\_ex[lp]}} p\lambda^{lp}_{(n,m)} - \sum_m p\lambda^{lp}_{(m,n)} = \begin{cases} 1, & \text{if } n = i\,(lp) \\ -1, & \text{if } n = j\,(lp) \\ 0, & \text{otherwise} \end{cases} \qquad \forall\ lp \in p\_LP;\ n \notin N\_ex[lp] \tag{15}$$

$$w\lambda^{lp}_{(m,n)} + p\lambda^{lp}_{(m,n)} \leq 1 \qquad \forall \, lp \in \text{p\_LP}; \; (m,n) \in N^2 \tag{16a}$$

$$\sum_{lp} w\lambda^{lp}_{(m,n)} + p\lambda^{lp}_{(m,n)} \leq W \qquad \forall \, (m,n) \in N^2 \tag{17a}$$

- With the integrated configuration approach:

$$\sum_{\substack{m \,:\, m \neq n, \\ m \notin N\_ex[lp]}} w\lambda^{(i,j),q}_{(n,m)} - \sum_{m} w\lambda^{(i,j),q}_{(m,n)} = \begin{cases} w\beta_{(i,j),q}, & \text{if } n = i \\ -w\beta_{(i,j),q}, & \text{if } n = j \\ 0, & \text{otherwise} \end{cases} \qquad \forall \, (i,j) \in N^2, \; n \notin N\_ex[lp], \; q \tag{14b}$$

$$w\lambda^{(i,j),q}_{(m,n)} + p\lambda^{lp}_{(m,n)} \leq 1 \qquad \forall \, (i,j),(m,n) \in N^2, \; q, \; lp \in \text{p\_LP}: i(lp) = i, j(lp) = j, q(lp) = q \tag{16b}$$

$$\sum_{(i,j)}\sum_{q} w\lambda^{(i,j),q}_{(m,n)} + \sum_{lp} p\lambda^{lp}_{(m,n)} \leq W \qquad \forall \, (m,n) \in N^2 \tag{17b}$$

4) Binary constraints:

$$w\beta_{(i,j),q}, p\beta_{(i,j),q}, w\delta^{lsp}_{(i,j),q}, p\delta^{lsp}_{(i,j),q}, w\lambda^{lp}_{(m,n)}, p\lambda^{lp}_{(m,n)}, w\lambda^{(i,j),q}_{(m,n)} \in \{0,1\} \tag{18}$$

Eq. (4) specifies the transit traffic at node $i$ as a difference between the total incoming traffic and the traffic terminated at $i$. Eq. (7), (8) make the number of lightpaths originating and terminating at node $i$ does not exceed T. For each LSP (working + protection), Eq. (9), (10) specify the flow conservation constraint at every node on its route. Some nodes are excluded from the routing of protection LSPs (Eq. (10)) to meet the protection routing constraints. Eq. (11) provides logical link disjointness between corresponding working and protection LSPs. Eq. (12) and (13) make the total traffic carried by all LSPs on the lightpath $(i,j)$, $q$ not to exceed the lightpath capacity. For each OXC and lightpath (working + protection), Eq. (14 a,b) and (15) specify the flow conservation constraint at the lightpath level. It states that the number of lightpaths incoming to and outgoing from a node is equal. With the integrated approach the logical topology and lightpath routing are determined jointly (Eq. (14 b)), whereas with the sequential approach lightpath routing is determined for the set of pre-computed lightpaths (Eq. 14a)). Eq. (16 a,b) provide link-disjointness between corresponding working and protection lightpaths. Eq. (17 a,b) make the number of lightpaths routed on the physical link $(m,n)$ not to exceed the total number of wavelengths. Binary constraint (18) ensures that the variables take only 0/1 values.

As the traffic matrix is symmetric and lightpaths bi-directional, for the problem simplification we assume that the lightpath set LP is computed only for the half of LSPs, i.e. $\text{LSP} = \{lsp_k(s,d,b): s < d\}$. It is assumed that the other half of LSPs: $\text{LSP}' = \{lsp'_k(s,d,b): s > d\}$ is routed over the corresponding complementary lightpaths. (i.e., for bi-directional lightpaths, if there exist the lightpath $lp_{(i,j),q}$ there exist also the lightpath $lp_{(j,i),q}$, so if the LSP $lsp_{k(s,d,b)}$ is routed onto $lp_{(i,j),q}$ than the $lsp'_{k(d,s,b)}$ is routed onto $lp_{(j,i),q}$).

In a network of $N$ nodes, $E$ links and max. $q$ lightpaths per node pair supporting a traffic matrix composed of $K$ LSPs, the size of the problem solved with the sequential approach is $\approx q \cdot N^2 \cdot K/2$ in terms of variables whereas the same problem solved with the integrated approach is $\approx q \cdot N^2 \cdot (K/2 + E)$.

V. NUMERICAL RESULTS

The problems specified above were implemented and solved using the CPLEX 9.0 optimization package. All experiments were carried out on a HP Alpha workstation with a 1 GHz CPU and 2 GB RAM running Tru64 UNIX OS. The system parameters were set as follows: C = 10 Gbps, W = 32, q = 2, T = $2q \cdot (N-1)$. The number of wavelengths per link W, IP/optical interface cards per node T, the router throughput and the OXC size were overprovisioned in a way that no resources shortage constraints were affecting the configuration process, but only the optimization function used.

## A. Computational efficiency

The logical topology design problem belongs to the class of NP-hard problems for which no efficient (i.e. polynomial time) algorithms are known. It is therefore essential to verify the computational limits of the proposed ILP-based solution method. The problem complexity measure was the execution time of the algorithms as a function of the problem size (in terms of the number of nodes and traffic demands). Solution times of the network configuration based on the sequential and integrated approaches were compared for different problem sizes and for the computation of working vs. protection paths. The solver was set to stop any optimization within a maximum time limit of 5 hours and a 3% solution optimality gap was assumed. If in any case the solver stopped without full termination, the solution quality achieved so far has been reported in terms of the optimality gap. The results are summarized in Table 1.

| Computation of working paths | | | | | | Computation of protection paths | | | | | |
|---|---|---|---|---|---|---|---|---|---|---|---|
| Config. Approach | Problem size (N) | | | | | Config. Approach | Problem size (N) | | | | |
|  | 8 | 10 | 12 | 14 | 15 |  | 8 | 10 | 12 | 14 | 15 |
| Sequential | 12.4 s (0.77%) | 70 s (0.72%) | 951 s (0.5%) | 1403 s (0.41%) | 3632 s (0.83%) | Sequential | 70 s (optimal) | 1011 s (0.44%) | 2542 s (0.66%) | 17093 s (0.32%) | unsolved |
| Integrated | 101 s (1.62 %) | 542 s (1.66%) | 1211 s (1.18%) | 18000 s (2.8%) | unsolved | Integrated | 550 s (0.34%) | 8856 s (1.12%) | 18000 s (1.76%) | unsolved | ----- |

Table 1. Running times of the ILP algorithms for different problem sizes (values in brackets denote achieved optimality gaps).

The obtained results show that by setting reasonable optimality gaps and run-time limits on ILP algorithms quite good solutions to the specified problems can be obtained even without a full termination, at least for moderate size networks (up to 15 nodes with fully meshed traffic matrices). Thereby, the proposed ILP-based solution method can be used as a practical design tool, for example, for nationwide backbone networks (which typically consist of 10-15 nodes in European countries) without resorting to heuristics. The solution times of the optimization based on the integrated approach is about one order of magnitude longer than when using the sequential approach and the difference in solution times for the two methods increases as the problem size grows. This was expected as in a network of $N$ nodes, $E$ links and max $q$ lightpaths per node pair the complexity of the integrated approach increases by $\sim q \cdot E \cdot N^2$ faster in terms of variables and constraints compared to the sequential approach. As a consequence, for the biggest solved problems the solution could only be obtained using the sequential approach.

Solution times for the computation of protection paths were longer due to additional constraints for protection routing: (11) and (15).

## B. Analysis of results

For the survivability implementation we used a test network targeted to a nationwide coverage, presented in Fig. 6. The network has a bi-connected mesh topology consisting of 12 nodes and 24 physical links. There are 10 nodes distributing the nationwide traffic (1-10) and 2 gateways providing the Internet access (11, 12). The traffic matrix is population-weighted and consists of 56 bidirectional, symmetric traffic demands or 126 LSPs (as some demands consist of more than 1 LSP). To check for different values of the LSP bandwidth and the total offered traffic, three groups of tests were carried out: for the average LSP bandwidth equal to 2.5, 5.0 and 7.5 Gbps. The results obtained with the sequential configuration method are summarized in Tables 2 a-c and Fig. 7.

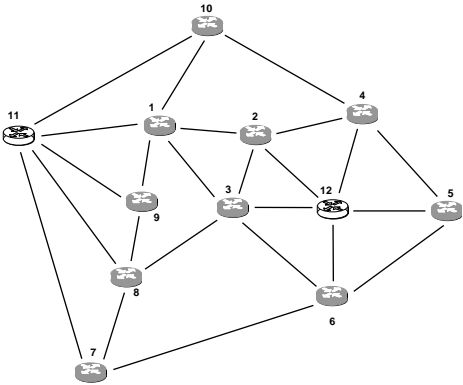

Fig. 6. The 12-node test network.

Table 2. Network resource usage and the configuration cost for different survivability options. The number of lightpaths in brackets (row 2) denotes the lightpaths carrying protection LSPs. The number of wavelengths in brackets (row 3) denotes extra wavelengths needed to accommodate protection LSPs with the "interlayer BRS". The cost percentage denotes the cost savings with respect to the most expensive method.

|  | Single layer Survivability | Multilayer Survivability | | |
|---|---|---|---|---|
|  |  | double protection | LSP spare unprotected | interlayer BRS |
| Transit traffic | 124 Gbps | 94 Gbps | 92 Gbps | 92 Gbps |
| No. of lightpaths | 56 (33) | 82 (18) | 66 (21) | 66 (21) |
| No. of wavelengths | 140 | 172 | 138 | 108 (4) |
| Total cost | 1471 (26%) | 1985 | 1626 (18%) | 1537 (22%) |
| Optical layer cost | 420 (19%) | 516 | 414 (25%) | 324 (37%) |

Tab. 2a. 126 LSPs; LSP bandwidth = 1, 2, 3 Gbps; total traffic = 300 Gbps

|  | Single layer Survivability | Multilayer Survivability | | |
|---|---|---|---|---|
|  |  | double protection | LSP spare unprotected | interlayer BRS |
| Transit traffic | 262.5 Gbps | 100 Gbps | 97.5 Gbps | 97.5 Gbps |
| No. of lightpaths | 143 (79) | 160 (16) | 148 (20) | 148 (20) |
| No. of wavelengths | 329 | 334 | 297 | 267 (7) |
| Total cost | 3628 (5%) | 3802 | 3485 (8%) | 3395 (11%) |
| Optical layer cost | 987 (1.5%) | 1002 | 891 (11%) | 801 (20%) |

Tab. 2b. 126 LSPs; LSP bandwidth = 2, 4, 6 Gbps; total traffic = 600 Gbps

|  | Single layer Survivability | Multilayer Survivability | | |
|---|---|---|---|---|
|  |  | double protection | LSP spare unprotected | interlayer BRS |
| Transit traffic | 107.5 Gbps | 52.5 Gbps | 52.5 Gbps | 52.5 Gbps |
| No. of lightpaths | 208 (105) | 226 (10) | 216 (10) | 216 (10) |
| No. of wavelengths | 596 | 505 | 490 | 480 (0) |
| Total cost | 5410 | 5399 (0.2%) | 5184 (4%) | 5154 (5%) |
| Optical layer cost | 1788 | 1515 (15%) | 1470 (18%) | 1440 (20%) |

Tab. 2c. 126 LSPs; LSP bandwidth = 3, 6, 9 Gbps; total traffic = 900 Gbps

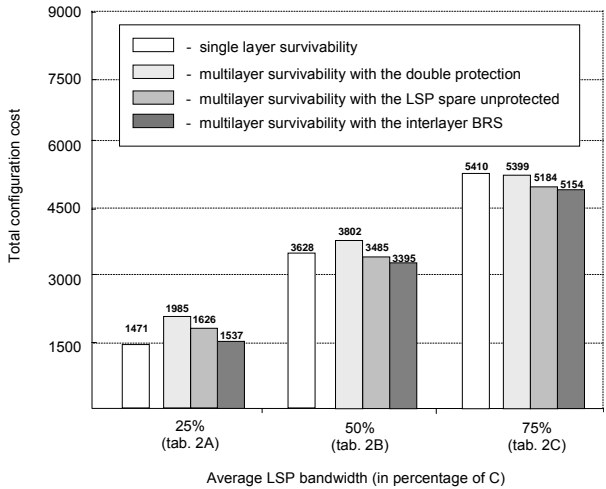

Fig. 7. Cost comparison of different survivability options.

*1) Single layer vs. multilayer survivability analysis.*

The main difference between the single layer (SL) and the multilayer (ML) survivability is due to the fact that with the ML survivability only multihop LSPs are subject to protection routing and hence consume the MPLS spare capacity, whereas with the SL survivability both single- and multi-hop LSPs require spare resources of the MPLS layer. This results in higher number of lightpaths carrying protection LSPs (see values in brackets in Tab. 2a-c) and higher transit traffic obtained with the SL survivability. However, with the ML survivability protection lightpaths are added, so the total number of lightpaths is higher in this scenario. This in turn affects the total configuration cost as lightpaths are the most expensive network resources.

For low bandwidth LSPs, the SL survivability is the cheapest option, as depicted in Tab. 2a and Fig. 7. Low bandwidth LSPs tend to be routed in multiple logical hops as they are groomed in intermediate routers to better fill the lightpath capacity. As a result, there exist relatively many multi-hop LSPs wich are subject to protection routing with the ML survivability and more lightpaths are added after the protection LSP routing (21), despite a small total number of lightpaths. This makes the ML survivability the most expensive option for low bandwidth LSPs.

This relation changes for high-bandwidth LSPs, close to the lightpath capacity (Tab. 2c). High bandwidth LSPs tend to be routed in single logical hops, i.e. on direct lightpaths, to minimize the transit traffic (note the small amount of the transit traffic, as compared with the other traffic scenarios). Therefore, there exist relatively few multi-hop LSPs which are subject to the protection routing with the ML survivability. As a result, much fewer lightpaths are added after the protection LSP routing (10), as compared with the SL survivability (105). This tendency and further savings in the wavelength use brought by the ML survivability (respectively 15%, 18% and 20% with different SCA methods) make this option the cheapest in this traffic scenario. The difference in the wavelength usage is due to the fact that with the SL survivability working and protection LSPs have disjoint routes in both network layers, whereas with the ML survivability working and protection lightpaths are disjoint only in the optical layer. As a result, lightpaths take longer routes with the SL survivability.

We found the ML survivability with the "double protection" the most expensive option for the low- and medium-size LSPs (see tab. 2a, b)

*2) Multilayer survivability – analysis of different SCA methods*

Decreasing cost trends from "double protection" to "LSP spare unprotected" to "interlayer BRS" was expected as the spare resources of the MPLS layer are supported more and more efficiently by the optical layer. Lower total number of lightpaths achieved with the "LSP spare unprotected" method over the "double protection" is due to the fact that the lightpaths carrying protection LSPs (respectively 21, 20 and 10 with different traffic scenarios) are not protected. Savings in the wavelength usage are because fewer lightpaths in total are routed in the physical topology (respectively 66 vs. 82, 148 vs. 160 and 216 vs. 226 for different traffic scenarios). Further savings are brought by the "interlayer BRS" method due to wavelength sharing among lightpaths carrying protection LSPs and protection lightpaths. One can see that most of the wavelengths used by the protection lightpaths are reused by the lightpaths carrying protection LSPs. Only, respectively, 4, 7, and 0 extra wavelengths are needed to accommodate protection LSPs within the optical layer for different traffic scenarios. This gives the reuse factor equal respectively to 84%, 80% and 100%.

*3) Sequential vs. integrated configuration method.*

As a next step we tested the impact of the sequential vs. integrated configuration method on network resource usage and configuration cost. The results produced by the two methods for different survivability options and the traffic matrix consisting of medium-size LSPs are summarized in Table 3 and Fig. 8. In all cases we observed a gain in the wavelength usage brought by the integrated method over the sequential one, while the MPLS layer resources (i.e. no. of established lightpaths and the transit traffic) were exactly the same. The difference is due to the fact that with the sequential approach the logical topology design and the lightpath routing are separated and the lightpaths are configured without the knowledge of the physical layer resources. As a consequence, the resulting logical topology is sub-optimal with respect to the wavelength usage and some lightpaths require longer physical routes. This effect is avoided with the integrated method where the both processes are configured jointly. This allows the lightpaths to be optimally designed also in terms of the wavelength use (in fact, about 40% of lightpaths had different termination points when configured with the integrated method as compared to the sequential one). The gain from using the integrated approach is higher for the configuration with working paths only (i.e. without survivability), as with the survivability implementation the wavelength routing is subject additionally to the protection routing constraints which tighten the solution space and leave less room for optimization.

| Survivability Option | Sequential | | | Integrated | | | Reduction | | |
|---|---|---|---|---|---|---|---|---|---|
| | Transit traffic (Gbps) | No. of lightpaths | No. of wavelengths | Transit traffic (Gbps) | No. of lightpaths | No. of wavelengths | Wavelengths | Optical layer cost | Total cost |
| No survivability | 77.5 | 64 | 108 | 77.5 | 64 | 93 | 15 | 14 % | 3 % |
| SL survivability | 262.5 | 143 | 329 | 262.5 | 143 | 303 | 26 | 8 % | 2 % |
| Double protection | 100 | 160 | 334 | 100 | 160 | 304 | 30 | 9 % | 2 % |
| LSP spare unprotected | 97.5 | 148 | 297 | 97.5 | 148 | 279 | 18 | 6 % | 1.5 % |
| Interlayer BRS | 97.5 | 148 | 267 | 97.5 | 148 | 251 | 16 | 6 % | 1.4 % |

Table 3. Sequential vs. integrated configuration method for different survivability options with the average LSP bandwidth equal to 50% of C (as in Tab. 2a).

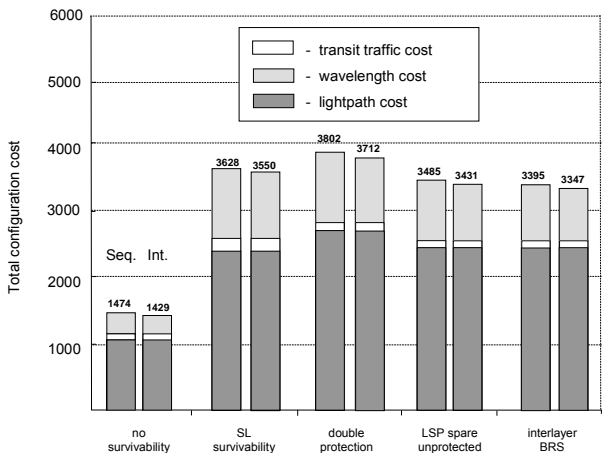

Fig. 8. Cost comparison of the sequential (Seq.) vs. integrated (Int.) configuration method.

As the number of lightpaths remains the same for both approaches which constitute about 70% of the total configuration cost (comp. Fig. 8), the contribution of the integrated approach to the total cost savings is only 1.4% - 3%.

VI. CONCLUSIONS

This study explored some design principles of MPLS over OTN architectures employing wavelength switching and targeted to a nationwide coverage. We used the ILP optimization combined with a cost model to dimension the network with the minimal resource usage and configuration cost. We have presented two approaches to the MPLS over OTN design and investigated various options for the survivability implementation.

The comparative analysis between the single- and multilayer survivability shows the impact of the LSP bandwidth on network resources usage and configuration cost. For high bandwidth LSPs, close to the lightpath capacity, the multilayer survivability is up to 5% more cost-effective in terms of the total cost and up to 20% in terms of the optical layer cost. Contrary, for the low bandwidth LSPs this is the single layer survivability which brings up to 26% of total cost savings. For the multilayer survivability we have demonstrated that by mapping efficiently the spare capacity of the MPLS layer onto the resources of the optical layer one can achieve up to 22% savings in the total configuration cost and up to 37% in the optical layer cost. We have found the integrated configuration method up to 9 % more cost-efficient in terms of the wavelength use as compared with the sequential one, however, at the increase in the optimization problem complexity. The obtained results are based on a cost model with actual technology pricing and representative traffic matrices.

Finally, we show that by setting reasonable optimality gaps and run-time limits on the ILP algorithms quite good solutions to the specified problems can obtained even without a full termination. Thereby, the proposed ILP-based solution method can serve as a practical design tool, at least for moderate size networks, without resorting to heuristics.